\documentstyle[aps,prl,twocolumn]{revtex}

\begin{document}

\draft

\title{
Analysis of self--averaging properties in the transport \\
of particles through random media
}

\author{J.M. L\'{o}pez, M.A.
Rodr\'{\i}guez and L. Pesquera }

\address{
 Departamento de F\'{\i}sica Moderna, Universidad de
Cantabria,
and Instituto
de Estudios Avanzados en F\'{\i}sica Moderna y Biolog\'{\i}a
Molecular,CSIC-UC,
Avenida Los Castros, E-39005 Santander, Spain.}

\maketitle

\begin{abstract}

We investigate self-averaging properties in the transport of particles
through random media. We show rigorously that in the subdiffusive anomalous
regime transport coefficients are not self--averaging quantities.
These quantities are exactly calculated in the case of directed
random walks. In the case of general symmetric random walks a perturbative
analysis around the Effective Medium Approximation (EMA) is
performed.

\end{abstract}


\narrowtext

The analysis of transport of particles in random media has interest in
physical systems since the transport mechanism is on the basis of many
physical phenomena \cite{r1,r2,r3,r4}.
The effect of disorder in the behavior of such systems can be normal, if
only a quantitative change of the transport parameters occur, or
anomalous when a qualitative change is induced by the disorder.
This anomalous behavior is of great interest in the
physics of disordered media and has been observed in almost all kind
of physical phenomena, from electrical conductivity to thermal properties
\cite{r2,r3,r4}.
There is no general theory for anomalous transport because in general
the involved phenomena are not unique. An important class of anomalous
behaviors are those due to restrictions in the motion of particles
imposed by the disorder. If this restriction is strong the motion of the
particle is subdiffusive and ,as a consequence, anomalies in the observed
phenomena occur.

An important characteristic associated to the anomalous behavior
is the sample to sample dependence of the measured quantities.
Strong sample to sample fluctuations are observed in most
cases of anomalous behavior \cite{r4}. In a normal situation the transport
coefficients are usually sample independent, while
for strong disorder transport coefficients are supposed not to be
self-averaging quantities.
The existence of sample to sample fluctuations is problematic either
from experimental or theoretical points of view. On the one hand,
experiments are usually performed over few samples. On the other
hand systematic analysis of disorder effects
have been usually based on the probability
density of diffusing particles averaged over random media configurations.
The study of  systems without self-averaging properties implies
the calculation of averaged products of probabilities or
some equivalent  function, which is a rather difficult task.
Only a few works have been devoted
to the analysis of self-averaging properties in special cases.
In \cite{derrida} a case of weak disorder was investigated, while
\cite{franceses1} and \cite{franceses2}
deal with cases of  directed random walks.
In recent works sample to sample fluctuations of first passage times
in asymmetric random walks \cite{ultimo1} have been investigated.
Also, and related with sample to sample fluctuations, there are
some works dealing with sample averaging of powers of the
probability \cite{prl}.

In this Letter we perform a
systematic analysis of the self-averaging properties of
systems  described by random walks (RW) on regular lattices
with a random distribution of transition rates. We introduce
a method based on the
renormalization of some coefficients of the evolution
equations to obtain averaged products of  probabilities.
This method generalizes the one previously used in \cite{nos} for the
calculation  of single averaged probabilities.
We apply the method, which we believe to be of rather wide
applicability, to the general symmetric RW and
to the directed RW in one dimension.
It is shown rigorously that for strong quenched disorder
the transport coefficients are not self-averaging.

We consider a general transport problem in which a particle moves in
a lattice with  random transition rates. The position of the
particle at a time t is denoted by $ r(t)$ and the magnitudes of
interest are mean functions of the position $F(r)$. Defining
$P(r,t)$ as the probability of finding the particle in $r$ at time
$t$, the observed quantities are given by:
\begin{equation}
F(t)=\sum_{r} F(r) P(r,t).
\label{1}
\end{equation}
In principle these quantities are dependent of the particular
configuration of the medium. A complete description of these
quantities can be achieved by using averaged moments of the form:
\begin{equation}
\langle F^n  (t) \rangle = \sum_{r_{1},r_{2}...r_{n}}
F(r_{1}) ...F(r_{n}) \langle P(r_{1},t)...P(r_{n},t) \rangle ,
\label{2}
\end{equation}
 where $\langle ... \rangle$ indicate average over
all possible configurations of the medium.
The self-averaging character of $F(t)$ can be derived from its variance.
A zero dispersion is equivalent to self-averaging. This dispersion can
be obtained from the calculation of the
averaged products of  probabilities. In the following we focus on this
problem in a transport model with a
probability governed by a master
equation with random coefficients  $w_{r}$ as:
\begin{equation}
 \partial _{t} P(R,t)= L_{0} P+L_{L} w_{r} L_{R} P ,
 \label{3}
 \end{equation}
where $L_{0}, L_{L} $ and $L_{R}$ are
linear operators. In a standard transport model these operators are
linear combinations of shift operators,
$L(r)= \sum_{i} a_{i} E_{i}(r) $, such that
$  E_{i} (r) P(r,t) = P(r+i,t) $
and $w_{r}$  are random transition rates.

The starting point of our method is the introduction of an
effective medium with memory \cite{nos}. In this way (\ref{3}) can be
written in terms of Laplace Transforms as
\begin{equation}
sP(r,s)-P_{0} (r)=[L_{0} + L_{L} \phi(s,r) L_{R}]P(r,s)
+ L_{L}[w_{r}-\phi(s,r)] L_{R} P(r,s),
\label{4}
\end{equation}
 where $P_0 (r) $ is the probability at $t=0$ and $\phi(r,s)$ is the transition
probability of the effective
medium that will be determined below. Taking  the integral
form of (\ref{4}) and iterating, a development in powers of the random
transition rate $\theta_{s}(r)=w_{r}-\phi(s,r)$ is obtained
\cite{nos}. Now we renormalize $\theta_{s}(r)$ by performing
a summation of all terms
in which contiguous indexes take the same value \cite{nos},
obtaining
\begin{equation}
P(r,s)=G_{s}(r, {r \prime}) P_{0}(r\prime)
+\sum_{n=1}^{\infty } \Phi_{s}({r_1 })..... \Phi_{s}({r_{n}})
G_{s}^{L}(r,r_{1}) J_{s}(r_{1},r_{2})....J_{s}(r_{n-1},r_{n})
G_{s}^{R}(r_{n},{r\prime}) P_{0}(r\prime),
\label{6}
\end{equation}
where summation over repeated indexes is understood
and  the sum is restricted to terms with
different contiguous indexes. This
renormalization corresponds to the one loop resummation
in diagrammatic representations, also known as single site
approximation in condensed matter.
The renormalized random transition is given by
\begin{equation}
\Phi_{s}(r)={\theta_{s}(r)\over 1-J_{s}(r,r)\theta_{s}(r)}
\label{7}
\end{equation}
and the functions $G_{s}^{L,R} $ and $J_{s}$  are defined by
\begin{equation}
G_{s}^{R} (r,r\prime)=L_{R}(r) G_{s}(r,r\prime) ,
G_{s}^{L} (r,r\prime)=L_{L}^{\dagger}(r\prime) G_{s}(r,r\prime)
\end{equation}
\begin{equation}
J_{s}(r,r\prime)=L_{R}(r) L_{L}^{\dagger}(r\prime) G_s(r,r\prime),
\end{equation}
 being $G_{s}(r,r')$ the propagator of the deterministic part of (\ref{4}),
$L_0 +L_L \phi L_R$.
Finally,  the transition probability
$\phi (r,s)$ is defined by the Effective Medium Approximation (EMA)
condition \cite{nos}: $\langle \Phi_{s} (r) \rangle = 0$.
When the model is translationally invariant the propagator is
only dependent on the difference of site positions and the effective
medium is homogeneous, that is, $\phi$ is not dependent on the position.

The  averaged products of  probabilities
can be directly calculated
from (\ref{6}). These products
are more conveniently
expressed in terms of $\delta P(r,s)$, defined as the difference
between the exact probability and that obtained with the effective
medium: $ \delta P(r,s)= P(r,s)- G_{s}(r,r \prime)
P_{0}(r\prime) $.
In this way the averaged products of $\delta P(r,t)$ are obtained
from (\ref{6}) as series in moments of $\Phi_{s} (r)$.
Since self-averaging is equivalent to a null dispersion, to analyze
the self-averaging character of the transport coefficients
only $\langle P(r, s)\rangle$ and $\langle P(r,s) P(r \prime ,s)\rangle$
must be considered. The method outlined above can be used to obtain
these magnitudes in a large variety of problems. Here we consider the
directed RW and the general symmetric RW in one-dimensional media with
quenched disorder.

\paragraph {Directed random walk (DRW) in 1D.}

In the DRW only steps in  one direction are allowed. Despite its simplicity
several phases or anomalous behaviors appear depending on
the intensity of disorder \cite{franceses1}. In one dimension
the master equation
modeling the DRW  can be written as:
\begin{equation}
\partial_{t} P(n,t)=-(1-E_{-1}(n)) w_n P(n,t) .
\end{equation}

The anomalous phases can be classified according to the intensity
of disorder, which is related to the existence of inverse moments
of the random term $w_{n}$.
If we restrict our analysis to the long time behavior of the velocity,
only the existence of
the first inverse moment is relevant.
Taking a probability
distribution $\rho (w_{n})=(1-\alpha) w_{n}^{- \alpha}$
the weak disordered phase correspond to the existence of the first inverse
moment, ($\alpha < 0$), and the strong disordered phase to
$\langle w_n ^{-1} \rangle  \to \infty$, ($1 > \alpha > 0$).
Other cases concerning
transients can be found in \cite{franceses1}.

The application of the method to this case is straightforward.
The propagator $G_s(n,m)$ is zero when $n <m$ and for $n \ge m$ we have
\begin{equation}
 G_{s}(n,m)={\phi (s)^{n-m} \over (s+\phi(s))^{n-m+1}} .
\end{equation}
Using the EMA condition we obtain the transition probability of the
effective medium $\phi (s) = R^{-1}(s) - s$
where the function $R(s)=\langle (s + w)^{-1} \rangle$ has been calculated
in Ref. \cite{franceses1}.
Since in the DRW only steps in one direction are possible,
 only terms  in (\ref{6})
with ordered indexes $r_1 > r_2 >...>r_n $ are different from zero.
Then $\langle \delta P(r,s) \rangle =0$ and $\langle P(r,s) \rangle$ is
exactly given by the EMA.
It is also possible to obtain exact expressions for the averaged
products of the moment generating function defined as
$F(x,s)=\sum _{i=0}^{\infty} x^{i} P(i,s) $.
The factorial moments $f_n (s)$ can be obtained by taking the derivative
of $F(x,s)$ at $x=1$.
All these quantities are sample dependent. The averaged products of
factorial moments can be calculated by means of averaged products of
generating functions as
\begin{equation}
\langle f_{n}(s) ... f_{m}(s)  \rangle={ { \partial^{n,...,m}
\langle F(x,s) ... F(y,s) \rangle }\over {\partial x^{n} ...
\partial y^{m} } } \bigg\vert_{x=...=y=1}.
\end{equation}

In general any self--averaging property can be analyzed with the
knowledge of the averaged products of generating functions.
Let us consider as an example the analysis of the behavior of the factorial
moments and assume the asymptotic form
$f_{n}(s) \sim a_{n} s^{\alpha _{n}} $, where $a_{n}$ is in principle
a sample dependent quantity.
The averaged value of $a_{n}$ and its dispersion
can be obtained from $\langle F(x,s) \rangle$ and from
$D(x,y,s)=\langle(F(x,s) -\langle F(x,s) \rangle)
(F(y,s) -\langle F(y,s) \rangle) \rangle $.
{}From (\ref{6}) we obtain the exact expression of the averaged products of
generating functions:
\begin{equation}
 \langle F(x,s) \rangle = {1\over s + \phi(s)(1-x)}
\label{F}
\end{equation}
\begin{equation}
D(x,y,s)={\langle \Phi_s^{2}\rangle \over (s+\phi(s))^{2}}
{(1-x) (1-y)\over\bigl[ s+(1-x)\phi(s) \bigr] \bigl[ s+(1-y)\phi(s) \bigr]}
{1 \over (1-A(s)xy) },
\label{D}
\end{equation}
where $A(s)= \bigl[\phi(s)^2\bigl(s+\phi(s)\bigr)^2
 s^2 \langle \Phi_s^2 \rangle \bigr] \bigl(s + \phi(s)\bigr)^{-4}$
and the renormalized random transition is
$ \Phi_s = \bigl(s+\phi(s)\bigr) \bigl[ \bigr(w-\phi(s)\bigr)/(w+s) \bigr] $.
{}From these expressions it is immediate the calculation of moments
and their sample to sample dispersions.
In the weak disordered phase, after calculation of $R(s)$ and
$\langle \Phi_s^{2} \rangle$, one obtains a ballistic behavior,
$\langle \overline r(t) \rangle \sim v t$, with a velocity $v= \langle
w^{-1} \rangle^{-1} $,
that is a self--averaging  quantity. In the strong disordered phase
one obtains, in agreement with [6], [7], a subballistic behavior,
$\langle \overline r(t) \rangle \sim b \, t^{(1-\alpha )}$
  with a coefficient with mean value $\langle b \rangle=
  \sin \bigl( \pi (1- \alpha) \bigr) / \bigl( \pi (1- \alpha )\Gamma
(2-\alpha) \bigl)$, that is not self--averaging.
The relative variance of $b$ is
  $\sigma_{r}^2 (b)= \langle (b-\langle b\rangle)^2 \rangle /\langle
b\rangle^2 =
\alpha / (2-\alpha)$. The dispersion increases
($\sigma_r (b) \to 1$) for stronger disorder ($\alpha \to 1$).

\paragraph {Symmetric random walk (SRW) in 1D.}

There are two models of  symmetric RW, the random trap (RT)
and random barrier (RB) models.
The master equations corresponding to both models are written in our
formulation as:
\begin{equation}
\partial_{t} P(n,t)=(1-E_{-1}(n)) w_{n}(E_{+1}(n)-1)P(n,t)
\end{equation}
for the random barrier and
\begin {equation}
\partial_{t} P(n,t)=(E_{-1}(n)+E_{+1}(n)-2) w_{n} P(n,t)
\end{equation}
for the random trap.
 The anomalous behavior induced by
the disorder is, in both cases, well known \cite{r1}.
The different phases can be classified following the definitions of [1].
We recall that for model A (weak disorder) the inverse moments of
$w_n$, $\beta_M
=\langle (w_n)^{-M} \rangle$ ($M=1,2,...$) are finite, while
 models B (marginal case, $\alpha=0$) and C (strong disorder, $0<\alpha
<1$)  are based on a probability distribution
$\rho (w_n )= (1-\alpha) \, w_n^{-\alpha}$ ($w_n \in
(0,1)$), such that inverse moments diverge.
In all cases the long time behavior of the sample averaged
diffusion coefficient has been exactly calculated. However sample
to sample fluctuations have not been investigated until now.
The sample averaged magnitudes are the same for RT and RB in one
dimension \cite{nos}. The same results are also obtained from (5) for the
self--averaging properties.

The application of the method to RB and RT models is also
straightforward, but it is
not possible to obtain exact expressions like in the DRW case.
The propagator $G_{s}(n,m)$ and
 the functions $J_{s}(n,m)$ are given in \cite{nos}
for all kinds of disorder.
The transition probability of the effective medium $\phi(s)$ has been
also calculated in \cite{nos} from the EMA condition. In
this reference we obtained the exact asymptotic behavior
of the averaged mean square displacement $\langle \overline
{x^{2}} (s) \rangle$ in the frequency domain, which is directly
related with the diffusivity.
The results derived from the EMA for each type of disorder
(A, B and C) are given by :
\begin{equation}
 \overline {x^{2}} (s)  _{EMA}=
  {2 \over s^2 \beta_{1}} \bigl( 1+{\beta_{2}-\beta_{1}^{2}
  \over 2 \beta_{1}^{2}}(\beta_{1}s)^{1\over 2} + {\rm O}(s) \bigr)
\end{equation}
\begin{equation}
 \overline {x^{2}} (s)  _{EMA}=
  {4 \over s^2 \mid \ln(s)\mid } \bigl( 1- {\ln\mid \ln(s) \mid
  \over \mid \ln s \mid } + {\rm O}(\mid \ln s  \mid^{-1}) \bigr)
\end{equation}
\begin{equation}
 \overline {x^{2}} (s)  _{EMA}=
2 \Bigl[ { {\sin (\pi \alpha)} \over { (1-\alpha)\pi 2^{\alpha}}  }
\Bigr]^{2/(2-\alpha)}
   s^{3 \alpha -4 \over 2- \alpha}  +
{\rm O}(s^{5 \alpha -6   \over 4-2\alpha} ).
\end{equation}

The exact results can be expressed as corrections to the
results given by the EMA as:
\begin{equation}
\langle \overline {x^{2}} (s) \rangle \sim
  \overline {x^{2}} (s)_{EMA} \bigl( 1+\alpha(s) \bigr)$$
\label{exact1}
\end{equation}
where the first corrections are
\begin{equation}
\alpha_{A} (s)= {1 \over 12 \beta _{1} ^{3}}(\beta_{2}-\beta_{1}^{2})^{2}
    s
\end{equation}
\begin{equation}
  \alpha_{B}(s)={(\pi^{2}+ 16 \ln 2- 20)\over \mid \ln s\mid^{2}}
\end{equation}
\begin{equation}
\alpha_{C}(s)= (4 \ln 2-5 +\pi^2 /4 ) \alpha^2 +
{\rm O}(\alpha^3 \ln \alpha).
\label{c}
\end{equation}
In the weak disordered case the behavior is normal and the EMA
reproduces exactly the first and second terms of
$\langle \overline{x^{2}} \rangle$.
In the marginal case B the EMA is exact up to terms
of order smaller than $ \mid \ln s \mid^{-3}$ \cite{nos}.
In the strong disordered case the behavior is subdiffusive and the EMA
does not reproduce exactly the coefficient of the leading term \cite{nos}.
Expressions (19)--(22)
have been diagrammatically calculated in \cite{nos} by using
cumulants and projection operators.
We can obtain the
same result from (\ref{6})  in a much more simple way
in terms of moments and single functions. This simplicity
allows us to calculate more involved quantities and to analize
self--averaging properties. For instance to analize the sample
to sample fluctuations of the generalized diffusion coefficient we
have calculated $\langle \overline{x^{2}}^{2} \rangle $ which depends
on the averaged product of probabilities. As in the above case
the exact result can be expressed as corrections to the EMA result
as
\begin{equation}
\langle \overline {x^{2}} (s)^{2} \rangle \sim
  \overline {x^{2}} (s)_{EMA}^{2} (1+\gamma(s) ),
\label{exact2}
\end{equation}
where the correction terms are
\begin{equation}
\gamma_{a}={\beta_{1}^{-{3\over2}}
(\beta_{2}-\beta_{1}^{2})\over 4} s^{1\over 2}
\end{equation}
\begin{equation}
\gamma_{b}={1\over \mid \ln s \mid}
\end{equation}
\begin{equation}
\gamma_{c}= \alpha /2 + O(\alpha ^2).
\end{equation}
 These terms have been calculated from a diagrammatic
representation of the averaged products of (\ref{6}). A detailed
description of this method will be
presented elsewhere.
Finally, from  (\ref{exact1}) and (\ref{exact2}) we can
extract several  conclusions. For weak and strong disorder the
dispersion of the particle can be taken, in the long time limit,
as $\overline{x^{2}} \sim a_{1} s^{\alpha_{1}}+a_{2} s^{\alpha_{2}} $,
where the coefficients are in principle sample dependent quantities.
In the weak disordered case the behavior is diffusive and we have
$\overline{x^{2}} \sim a_{1} s^{-2}+a_{2} s^{-7/4} $, where the
first coefficient  is self--averaging but the second is sample dependent
with a zero mean value and a variance $\sigma^2 (a_2) = \beta_1^{-7/2}
(\beta_2 -\beta_1^{2})$.
In the strong disordered case the behavior is subdiffusive,
$\overline{x^{2}} \sim c_{1} s^{-(4- 3\alpha)/(2-\alpha)} $ and $c_1$
is not self--averaging.
The mean value of the coefficient $c_1$ can be easily calculated for
small $\alpha$ from (\ref{exact1}) and (\ref{c})
($\langle c_1 \rangle = 4\alpha + O(\alpha^3)$).
As we will show elsewhere its dispersion
can be also obtained in the same way $\sigma^2(c_1) = 2 \alpha^3
+ O(\alpha^4 \ln \alpha)$. Finally, the marginal
case B is similar to the weak case with logarithmic corrections and we have
$\overline{x^{2}} \sim b_{1} s^{-2} \mid \ln s\mid^{-1} +b_{2}
s^{-2} \mid \ln s \mid^{-3/2}  $. The first coefficient is
self-averaging but $b_2$ is sample dependent with a zero mean value and
a variance $\sigma^2 (b_2) = 16$.

In summary, we have presented here a general method to study self--averaging
properties in the transport on random media. Our analysis of both DRW and SRW
shows rigorously that when the behavior of a magnitude is normal its
 long time behavior is
sample independent. By the contrary in anomalous diffusion phases
the  self--averaging property is not satisfied.
The method introduced in this letter can be easily applied to other
one dimensional problems and it can be also extended to more dimensions.
Some of these applications will be presented elsewhere.

Financial support from Direcci\'on General de Investigaci\'on Cient\'\i fica
y T\'ecnica (Spain), Project No. PB93-0054-C02-02 is acknowledged.

\end{document}